\documentclass{optica-article}

\journal{opticajournal} 

\articletype{Research Article}

\usepackage{lineno}
\usepackage{bm}
\usepackage{chemformula}
\usepackage{upgreek}

\begin{document}

\title{A spin-embedded diamond optomechanical resonator with  mechanical quality factor exceeding one million}

\author{Hyunseok Oh,\authormark{1,*} Viraj Dharod,\authormark{1,*} Carl Padgett,\authormark{1,*} Lillian B. Hughes,\authormark{2} Jayameenakshi Venkatraman, \authormark{1} Shreyas Parthasarathy, \authormark{1} Ekaterina Osipova,\authormark{1} Ian Hedgepeth,\authormark{1} Jeffrey V. Cady,\authormark{1}  Luca Basso,\authormark{3}, Yongqiang Wang,\authormark{4} Michael Titze,\authormark{5} Edward S. Bielejec,\authormark{5}, Andrew M. Mounce,\authormark{3}, Dirk Bouwmeester,\authormark{1,6} and Ania C. Bleszynski Jayich\authormark{1,†}
}

\address{\authormark{1}Department of Physics, University of California, Santa Barbara, California 93106, USA\\
\authormark{2}Materials Department, University of California, Santa Barbara, California 93106, USA\\
\authormark{3}Center for Integrated Nanotechnologies, Sandia National Laboratories, Albuquerque, New Mexico 87123, USA\\
\authormark{4}Center for Integrated Nanotechnologies, Los Alamos National Laboratory, Los Alamos, New Mexico 87545, USA\\
\authormark{5}Sandia National Laboratories, Albuquerque, New Mexico 87185, USA\\
\authormark{6}Huygens-Kamerlingh Onnes Laboratory, Leiden University, P.O. Box 9504, 2300 RA Leiden, Netherlands\\
\authormark{*}These authors contributed equally to this work.
}

\email{\authormark{†}ania@physics.ucsb.edu} 


\begin{abstract*} 
Diamond optomechanical crystal (OMC) devices with embedded color center spins are promising platforms for a broad range of applications in quantum sensing, networking, and computing applications, offering an interface between a GHz-frequency mechanical mode and both optical photons and coherent spins.
A crucial but elusive step towards realizing this platform is to engineer a device with a high-quality factor mechanical mode while preserving the bulk-like coherence of embedded spins.
Here we demonstrate sideband-resolved diamond OMCs with mechanical quality factors in excess of $10^6$ at cryogenic temperatures, and find coherence times up to $T_2$ = 270 $\upmu$s for embedded nitrogen vacancy (NV) centers.
Furthermore, we measure these devices across five orders of magnitude in intracavity optical power, demonstrating robust power handling and a high optomechanical cooperativity ($C\gg1$) at cryogenic temperatures that is essential for a broad range of quantum protocols requiring strong, coherent interactions between photons and phonons.
These results are enabled by a robust, high-throughput method for forming single-crystal diamond membranes in combination with chemical vapor deposition (CVD) diamond overgrowth with nitrogen $\delta$-doping.
We discuss the prospects of this platform for hybrid spin-mechanical devices in the quantum regime.
\end{abstract*}

\section{Introduction}
Mechanical systems have risen to prominence in the fields of quantum sensing and quantum information science due to their long second-scale lifetimes \cite{maccabe_nano-acoustic_2020} and demonstrated operation in the single-phonon quantum regime \cite{wallucks_quantum_2020, wollack_quantum_2022}.
They can serve as key auxiliary components in hybrid quantum systems, offering versatile coupling to diverse quantum degrees of freedom, including photons \cite{cohadon_cooling_1999}, charge qubits \cite{lahaye_nanomechanical_2009}, spin qubits \cite{rugar_single_2004}, and solid-state spins \cite{macquarrie_mechanical_2013}, while maintaining high quality factors.
For example, they can facilitate coupling between distant quantum systems \cite{lee_topical_2017} and enable quantum information transfer across different energy scales \cite{andrews_bidirectional_2014,mirhosseini_superconducting_2020}.
Mechanical systems have great potential for quantum sensing \cite{purdy_strong_2013,kronwald_arbitrarily_2013,wollman_quantum_2015}, quantum memories \cite{wallucks_quantum_2020}, and quantum transduction \cite{mirhosseini_superconducting_2020,bienfait_phonon-mediated_2019}.

Single-crystal diamond is a promising material platform for hybrid mechanical systems, featuring excellent optical, mechanical, and thermal properties \cite{coe_optical_2000}.
Importantly, it hosts a variety of highly coherent defect-center spins, such as nitrogen-vacancy (NV) and silicon-vacancy (SiV) centers \cite{lee_topical_2017}.
Theoretical proposals suggest the use of an engineered coupling between these localized spin-based qubits and the long-lived delocalized phononic modes of diamond mechanical oscillators for, \textit{e.g.}, quantum networking, improved spin readout, and mechanically mediated spin-spin entanglement \cite{koppenhofer_single-spin_2023,koppenhofer_dissipative_2022,groszkowski_heisenberg-limited_2020}. 
Several experiments have demonstrated strain-mediated coupling between diamond mechanical modes and embedded NV \cite{ovartchaiyapong_dynamic_2014,lee_strain_2016} and SiV centers \cite{meesala_strain_2018, maity_coherent_2020,joe_observation_2025}.

In optomechanical devices, optical control over mechanical modes can enable, for instance, ground state cooling \cite{teufel_sideband_2011, chan_laser_2011} and remote quantum entanglement of mechanical oscillators \cite{riedinger_remote_2018}. 
A key figure of merit for accessing the quantum regime of mechanical motion in these devices is the optomechanical cooperativity, $C = \frac{4g_0^2n_\mathrm{c}}{\kappa\gamma_\mathrm{m}}$, where $g_0$ represents the optomechanical coupling strength (\textit{i.e.}, the energy exchange rate between photons and phonons), $\kappa$ and $\gamma_\mathrm{m}$ respectively denote the energy loss rates of the optical and mechanical modes, and $n_\mathrm{c}$ is the average number of photons in the optical cavity.
In cavity optomechanics, efficient cooling to the ground state is possible with large cooperativity and in the resolved-sideband regime  \cite{schliesser_resolved-sideband_2008, teufel_sideband_2011, chan_laser_2011}, where the mechanical resonance frequency $\omega_\mathrm{m}$ exceeds the optical cavity linewidth $\kappa$ \cite{marquardt_quantum_2007}.

Optomechanical crystals (OMCs), comprising photonic and phononic crystal structures on the nano- to micro-meter scale \cite{eichenfield_optomechanical_2009}, host co-localized optical and mechanical modes that can be engineered with high quality factors.
Due to their high mechanical frequency and amenability to optomechanical quantum control, they serve as an excellent starting point for operation in the quantum regime.
Ground state cooling \cite{chan_laser_2011} and entanglement of mechanical quantum states \cite{riedinger_remote_2018} have been demonstrated with GHz-frequency silicon OMCs.

The attractive features of OMCs combined with the prospect of strain-mediated coupling to embedded spins in defect-containing hosts provide strong motivation to realize spin-coupled diamond OMC devices.
In recent years, there has been progress in fabricating diamond OMCs \cite{burek_diamond_2016, cady_diamond_2019, joe_high_2024,joe_observation_2025}, though experiments are still sparse due to challenges in nanofabrication, and the demonstrated optomechanical properties have lagged their Si counterparts.
One major challenge is the formation of thin films of single-crystal diamond, and several approaches have been pursued to this end.
In the diamond-on-insulator (DOI) method \cite{ovartchaiyapong_high_2012, tao_single-crystal_2014}, diamond membranes are bonded to another material and subsequently thinned to reach the target device thickness.
However, the resulting membranes typically exhibit thickness variations of several microns across a mm-scale lateral extent, severely limiting the throughput of devices with a target thickness.
Other techniques include top-down fabrication of bulk diamond such as angled etching \cite{burek_diamond_2016, joe_high_2024} and quasi-isotropic etching \cite{khanaliloo_high-qv_2015,mitchell_realizing_2019,joe_observation_2025}.

Another promising method for forming high-quality, uniform-thickness, single-crystal diamond thin films for spin-coupled diamond optomechanics involves the smart-cut technique \cite{bruel_smart_1995}, a well-established nanofabrication technique for materials beyond diamond, followed by subsequent chemical vapor deposition (CVD) diamond overgrowth \cite{ohno_engineering_2012}.
The diamond smart-cut process creates a subsurface damaged layer that can be selectively removed, thus enabling the separation of a uniformly thin diamond layer from the bulk.
This method has been explored for diamond optical and mechanical nanostructures \cite{fairchild_fabrication_2008,lee_fabrication_2013,piracha_scalable_2016,guo_tunable_2021, guo_direct-bonded_2024, ding_high-q_2024, basso_fabrication_2024} and features several advantages. 
The thickness uniformity enables high throughput fabrication of nanostructures across the full diamond membrane.
The rectangular cross-section, in contrast with the triangular cross-section from angled-etching approaches, eliminates the need for specialized etching techniques.   
Furthermore, a rectangular cross-section easily enables acoustic shielding structures around the OMC that are required for a full phononic band gap \cite{chan_optimized_2012}.
The mechanical breathing mode induces a uniform strain profile along the nanobeam thickness, permitting spins to be placed at the midpoint in depth where they are protected from surface-induced decoherence \cite{myers_probing_2014} and allowing for a larger uniform strain region. 
We note that the subsequent CVD overgrowth of high quality diamond on smart-cut membranes is essential for good spin, optical, and mechanical properties \cite{lee_deterministic_2014}.

Here we present diamond OMC devices fabricated via the smart-cut method with high mechanical quality factors $Q_\mathrm{m} = (1.90 \pm 0.04) \times 10^6$ and optical quality factors $Q_\mathrm{o} = (3.781 \pm 0.004) \times 10^4$ in the resolved-sideband regime. The frequency-$Q$ product for the 6.23 GHz mechanical mode is $1.18 \times 10^{16}$ Hz. 
We characterize the optomechanical properties at liquid-He temperatures, demonstrating an optomechanical cooperativity of 54 measured at a large intracavity photon number $n_\mathrm{c} = 41,000$ enabled by the high power handling capability of diamond. 
Additionally, we measure the room temperature spin properties of an NV center embedded within the OMC and observe a long spin coherence time of 227 $\upmu$s as well as high spin-dependent optical contrast.
We discuss the potential for strong mechanical coupling of these devices to embedded spins in the quantum regime.

\begin{figure}[!htb]
\centering\includegraphics[width=1\textwidth]{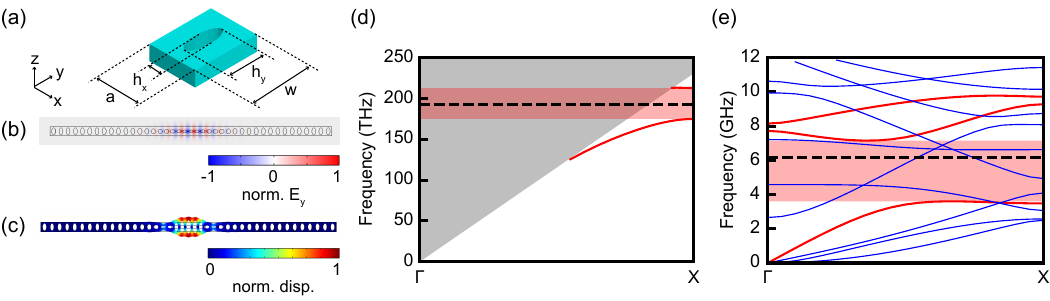}
\caption{\label{fig:intro}
Design and simulation of 1D diamond optomechanical crystal (OMC).
(a) Schematic of an OMC unit mirror cell. The indicated dimensions are varied through optimization with final values $(a, w, h_x, h_y) = (650, 800, 343, 617)$ nm and nanobeam thickness $t = 250$ nm.
(b) Simulated fundamental TE-like optical mode profile of the cavity resonance at $\omega_\mathrm{o}/2\pi$ = 192 THz. $E_y$ is the normalized y-component of the electric field. 
(c) Simulated fundamental mechanical breathing mode profile of the cavity resonance at $\omega_\mathrm{m}/2\pi$ = 6.14 GHz. The color represents the normalized displacement field amplitude.
(d) Simulated optical band structure of the unit mirror cell. The red lines correspond to the optical modes of both odd vector symmetry in the y-axis and even vector symmetry in the z-axis. The grey-shaded region denotes the light cone with unguided modes.
(e) Simulated mechanical band structure of the unit mirror cell. The red lines are mechanical modes of even vector symmetry in both the y and z-axis while the blue lines correspond to other modes. 
}
\end{figure}

\section{Device design and fabrication}
Our photonic/phononic cavity structure is formed by patterning elliptical holes of varying size and spacing to form cells (Fig. \ref{fig:intro}a) along the length of a 1D nanobeam.
The central defect cell hosts the fundamental transverse-electric-like (TE) optical and mechanical breathing modes, while mirror cells on either side of the nanobeam host symmetry-dependent optical and mechanical quasi-bandgaps to confine the mechanical and optical modes of interest.
Cells between the defect cell and the mirror cells adiabatically transition the cell geometry to minimize scattering losses.
The optical and mechanical modes of our geometry are simulated using the finite element method software COMSOL, and the resulting mode profiles are shown in Fig. \ref{fig:intro}b, c.
The photonic and phononic band structure diagrams of the mirror unit cells are shown in Fig. \ref{fig:intro}d, e, where red-shaded regions indicate quasi-bandgaps surrounding the frequencies of the optical and mechanical modes of interest, indicated by dashed black lines.

The dimensions of the OMC device are chosen to optimize the optomechanical coupling strength ($g_0$) and the optical quality factor ($Q_\mathrm{o}$), while maintaining a high mechanical quality factor.
We applied a genetic algorithm to explore the design parameter space and maximize the product $g_0Q_\mathrm{o}$, where $g_0$ and $Q_\mathrm{o}$ are extracted from COMSOL simulations. 
The parameters are constrained to ensure the mechanical and optical modes lie within their respective quasi-bandgaps.
The simulations of the optimized OMC device geometry yield an optical resonance at $\lambda_\mathrm{o} = 1559$ nm, a mechanical resonance at $\omega_\mathrm{m} /2\pi = 6.14$ GHz, and $g_0/2\pi = 201$ kHz.
Detailed optimization procedures are described in Section 3 of the Supplementary Material.

\begin{figure}[!htb]
\centering\includegraphics[width=0.50\textwidth]{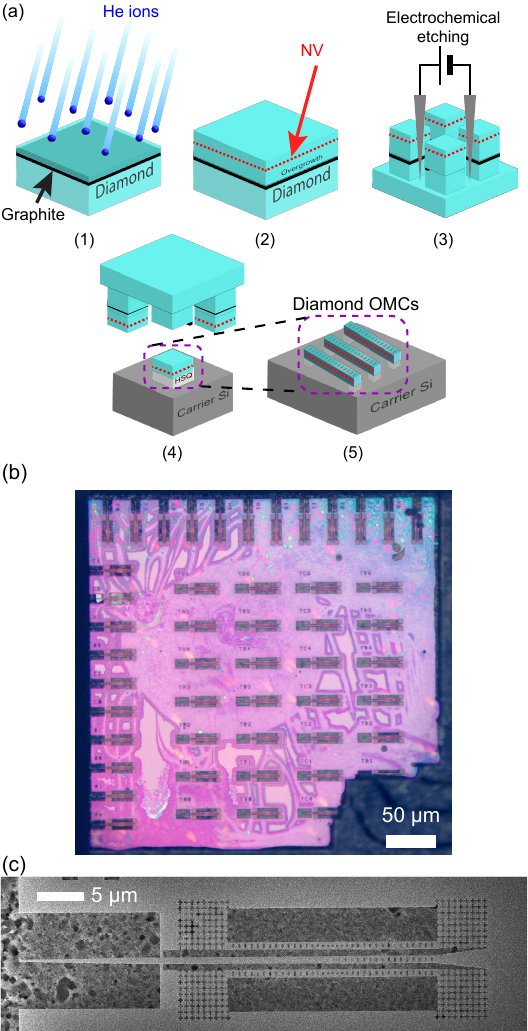}
\caption{\label{fig:fab} 
(a) Steps for diamond OMC device fabrication: (1) Sub-surface damage layer formation, (2) CVD diamond overgrowth with \ch{^{15}N} $\delta$-doping, (3) Electrochemical etching of patterned membranes, (4) Membrane transfer to a Si carrier piece, and (5) Patterning OMC devices. 
(b) Optical microscope image of a diamond membrane with approximately 50 fabricated OMC devices.
(c) SEM image of a device, showing a pair of OMCs surrounded by acoustic shields and a tapered waveguide for optical coupling. 
}
\end{figure}

Fig. \ref{fig:fab}a summarizes the steps for fabricating OMCs in thin-film diamond.
We start with an electronic grade 100-oriented single-crystal diamond sample (Element Six) polished to a surface roughness ($R_q$) of $<$ 300 pm.
A subsurface damaged layer is formed by He ion implantation at an energy of 150 keV and a fluence of $5\times10^{16}$ ions / cm$^2$, followed by graphitization upon high-temperature vacuum annealing. 
The result is a 400 nm-thick diamond layer atop a 100 nm-thick graphite layer. 
The top diamond layer is partially damaged due to the traversal of He ions, and hence to improve the device material quality, we use CVD to grow a 497 nm-thick $^{12}$C isotopically purified layer with a delta-doped \ch{^{15}N} layer at a depth of 144 nm.
After growth, NV centers are formed via 200 keV electron irradiation and 850 °C annealing for 8 hours.
We next pattern 16 square membranes (400 $\upmu$m $\times$ 400 $\upmu$m) on the $2 \times 2$ mm diamond via photolithography and plasma etching. 
We electrochemically etch the graphite below a single target membrane in 250 mM \ch{K2SO4} aqueous solution with Pt wires \cite{tully_diamond_2021} until only a small corner region of graphite tethers the diamond membrane to the substrate.
Next the diamond membrane is transferred via flip chip bonding onto a Si carrier piece spin-coated with hydrogen silsesquioxane (HSQ) and heated to 450 °C to cure the HSQ under 400 N of applied force.
After HSQ curing, the bulk diamond sample is retracted from the silicon, leaving only the target membrane bonded to the Si piece.
We then use ArCl$_2$ and O$_2$ reactive ion etching (RIE) to thin the membrane down to the target thickness of 250 nm, removing the partially He-ion-damaged diamond in the process.

We fabricate approximately 50 OMC devices on a transferred membrane (Fig. \ref{fig:fab}b) via e-beam lithography (EBL), followed by O$_2$ RIE. 
Each device (Fig. \ref{fig:fab}c) consists of an OMC nanobeam cavity, an optical waveguide for delivering light from an optical fiber to the device, and acoustic shields for suppressing the clamping-induced mechanical damping.
The devices are diced within a few $\upmu$m of the diamond device edge to enable optical coupling to a lensed fiber with a 14 $\upmu$m working distance.
Detailed fabrication procedures are described in Section 1 of the Supplementary Material.
We note that the production rate for diamond OMCs using this membrane technique is significantly higher than our previously used diamond-on-insulator (DOI) approach \cite{cady_diamond_2019}.
Many useful devices can now be fabricated on one membrane (see Section 4D of the Supplementary Material).

\section{Optomechanics characterization}
\begin{figure}[!htb]
\centering\includegraphics[width=0.50\textwidth]{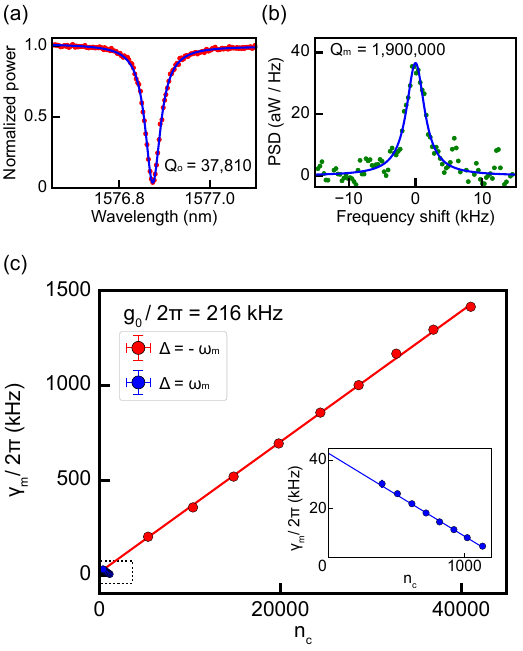}
\caption{\label{fig:opto} 
(a) Optical resonance of the diamond OMC. A Fano-resonance fit (blue line) gives an optical quality factor of $(3.781 \pm 0.004) \times 10^4$.
(b) Mechanical resonance of the diamond OMC at T = 4 K. A Lorentzian fit (blue line) gives a mechanical quality factor of $(1.90 \pm 0.04) \times 10^6$ at a center frequency of 6.23 GHz.
(c) Total mechanical linewidth ($\gamma_\mathrm{m}$) as a function of intracavity photon number ($n_\mathrm{c}$) due to backaction cooling (laser on red motional side band) and amplification (laser on blue motional side band).
The data is collected at 4 K.
Fitting the lines to Eq. \eqref{eqn:g} yields an optomechanical coupling strength ($g_0$) of $216 \pm 1$ kHz. Inset: Magnified blue-sideband data at low $n_\mathrm{c}$.
}
\end{figure}

We first present optical and mechanical spectroscopy of a diamond OMC device at cryogenic temperatures.
We sweep the wavelength of a tunable external cavity diode laser to measure the optical response of the OMC (Fig. \ref{fig:opto}a) and extract an optical resonance wavelength of $\lambda_\mathrm{o} = 1576.87$ nm and an optical quality factor of $Q_\mathrm{o} = (3.781 \pm 0.004) \times 10^4$.
With the laser tuned to the red motional sideband of the OMC cavity, the mechanical Brownian motion of the cavity imprints a modulation at the mechanical frequency via optomechanical coupling.
The power spectral density of the modulated tone for an intracavity photon number of $n_\mathrm{c} = 0.27$ is shown in Fig. \ref{fig:opto}b, from which we extract a mechanical quality factor of $Q_\mathrm{m} = (1.90 \pm  0.04) \times 10^{6}$ at a sample stage temperature of 4 K. The mechanical resonance frequency is $\omega_\mathrm{m}/2\pi = 6.23$ GHz. 
Measurements performed at a sample stage temperature of 160 mK yield a comparable, though slightly lower, $Q_\mathrm{m}$ (see Section 4C of the Supplementary Material).
These measured parameters put our diamond OMCs in the resolved-sideband regime, with $4\omega_\mathrm{m} / \kappa=4.96$.
We note that this is the highest $Q_\mathrm{m}$ reported for diamond OMC devices to date\cite{joe_high_2024} and it is higher than the frequency jitter-limited $Q_\mathrm{m} \approx 1.31 \times 10^{6}$ of Si OMCs reported by MacCabe \textit{et al.} \cite{maccabe_nano-acoustic_2020}, where significantly higher $Q_\mathrm{m}$ was also measured for the same device via ringdown measurements with short optical pulses.
Future experiments will probe diamond OMCs using pulsed measurement schemes and will elucidate the origins of mechanical damping in these devices. 

We characterize the optomechanical coupling between the optical and the mechanical modes by measuring the backaction-induced mechanical damping rate as a function of $n_\mathrm{c}$.
With a laser tuned to the red ($\Delta = -\omega_\mathrm{m}$) or blue ($\Delta = \omega_\mathrm{m}$) motional sideband, the total mechanical damping rate is given by 
\begin{equation}\label{eqn:g}
    \gamma_\mathrm{m} = \gamma_\mathrm{i} \mp \frac{4g_0^2 n_\mathrm{c}}{\kappa} \left(\frac{ \left(\frac{4\omega_\mathrm{m}}{\kappa}\right)^2}{1+\left(\frac{4\omega_\mathrm{m}}{\kappa}\right)^2} \right),~ \Delta = \pm \omega_\mathrm{m},
\end{equation}
where $\gamma_\mathrm{m}$ is the total mechanical damping rate, $\gamma_\mathrm{i}$ is the intrinsic mechanical damping rate, $g_0$ is the optomechanical coupling strength, $n_\mathrm{c}$ is the average number of photons in the cavity, and $\kappa$ is the optical decay rate \cite{aspelmeyer_cavity_2014}.
A fit to data collected at T = 4 K (Fig. \ref{fig:opto}c) yields $g_0/2\pi = 214.8 \pm 0.8$ kHz on the red-sideband and $g_0/2\pi = 218 \pm 3$ kHz on the blue-sideband. 
At the highest measured $n_\mathrm{c}$ of 41,000, our device achieves an optomechanical cooperativity $C = 54 \pm 5$, demonstrating robust optical power handling. 
For comparison, previously reported $C$ values for diamond OMCs range up to 19.9 \cite{burek_diamond_2016}. 
We note that this cooperativity is calculated using $\gamma_\mathrm{i}/2\pi$ = 28 kHz extracted from fitting the data in Fig. \ref{fig:opto}c to Eq. \ref{eqn:g}.
The 3.28 kHz damping rate measured in Fig. \ref{fig:opto}b is lower presumably due to decreased parasitic optical absorption at much lower $n_\mathrm{c}$.
The $n_\mathrm{c}$ and $T$ dependence of $Q_\mathrm{m}$ will be the subject of future study.
In this regime of $C\gg1$, photons and phonons interact coherently, making diamond OMCs a promising candidate platform for \textit{e.g.} ground state cooling, quantum state transduction, and optomechanically-assisted single-spin readout \cite{koppenhofer_single-spin_2023}.

\section{NV center characterization}
With the goal of integrating defect-based spin qubits into optomechanical structures, it is important that the fabrication process preserves the qubit properties, leaving them as `bulk-like' as possible.
The NV center is an optically-addressable spin in diamond exhibiting long room-temperature quantum coherence time ($T_2$), and coupling its spin degree of freedom to nanoscale mechanical and optical cavities is promising for a variety of sensing and networking applications \cite{jensen_cavity-enhanced_2014, kaupp_purcell-enhanced_2016}.
However, the NV center's spin and optical properties are known to be highly sensitive to nearby surfaces and proximal lattice damage induced by nanofabrication \cite{rodgers_materials_2021}.

\begin{figure}[!htb]
\centering
\includegraphics[width=0.5\textwidth]{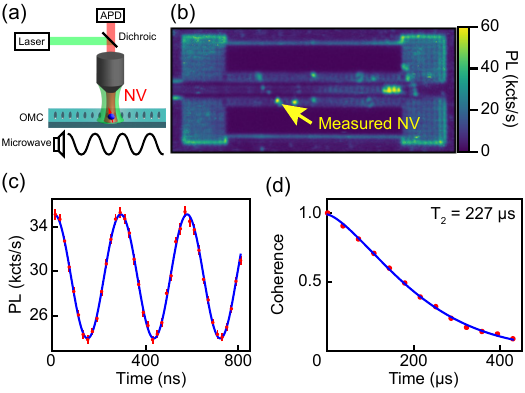}
\caption{\label{fig:nv} 
Spin properties of NV centers embedded in OMCs.
(a) Confocal microscopy setup.
(b) Scanning confocal microscope image of an OMC device.
(c) Rabi oscillations of the NV spin indicated by the yellow arrow in (b), where the NV photoluminescence (PL) oscillates as a function of the duration of an applied microwave pulse. Fitting the visibility of the Rabi fringes (the Rabi contrast) gives 31.2 $\pm$ 0.2 \% Rabi contrast.
(d) Hahn-echo coherence decay curve for the NV indicated in (b). 
A stretched exponential decay fit (stretch exponent $n$ = 1.44 $\pm$ 0.04) yields $T_{2}$ = 227 $\pm$ 3 $\upmu$s.
The coherence was sampled at times matched to the period of \ch{^{15}N} electron spin echo envelope modulation (ESEEM) to trace the decay envelope \cite{ohno_engineering_2012}.
}
\end{figure}

Here we investigate NV centers inside fabricated OMCs to assess the compatibility of our fabrication process with the preservation of high quality spin properties. 
We use a room temperature confocal microscope (Fig. \ref{fig:nv}a) to excite the NV centers with a 532 nm laser, and collect fluorescence in the 650-800 nm range with an avalanche photodiode (APD).
Microwaves for driving NV spin transitions are delivered via an off-chip microwire. 
Fig. \ref{fig:nv}b shows a scanning confocal image of one of our devices, where many of the bright spots correspond to fluorescent NV centers, as verified by optically detected magnetic resonance (ODMR) spectroscopy.
Fig. \ref{fig:nv}c shows Rabi oscillations of the NV spin qubit encoded in the $m_s = 0$ and $m_s = -1$ spin states; the data shows an excellent spin-dependent fluorescence contrast of $\sim$ 31\%, which is close to the maximum Rabi contrast typically seen for charge-stable NV centers \cite{yuan_charge_2020}.
The Rabi contrast is defined as $(PL_0-PL_{-1})/PL_0$ where $PL_0$ ($PL_{-1}$) is the photoluminescence rate of the NV in the $m_s=0$ ($m_s=-1$) state.
Both the fluorescence rate and high contrast persist over prolonged laser illumination, indicating that the good NV charge state initialization fidelity is stable over time and robust to high light intensity.
In Fig. \ref{fig:nv}d, we plot the results of a Hahn-echo coherence time measurement and fit the data to extract $T_2$ = $227 \pm 3 \,\upmu$s. 
These results confirm the gentle nature of our diamond growth, defect incorporation, and device nanofabrication process, which preserves the charge stability and spin coherence of embedded NV centers.

\section{Conclusion and outlook}
The diamond OMC devices presented here exhibit high mechanical quality factors exceeding 1.9 million and optical quality factors that place them in the resolved-sideband regime.
These results are enabled by a diamond smart-cut technique combined with CVD diamond overgrowth to fabricate uniformly thin nanoscale diamond membranes of high material quality.
The high optomechanical cooperativity of 54 at a circulating photon number of 41,000 demonstrates a high tolerance for intracavity optical power that is promising for experiments where strong and coherent interactions between photons and phonons are required.
Our devices compare favorably to state-of-the-art 1D Si OMC devices in terms of mechanical quality factors measured under continuous-wave (CW) optical probing (see Section 4E of the Supplementary Material) and cooperativity, for which good power handling compensates relatively lower optomechanical coupling due to diamond’s lower refractive index.
We have also demonstrated optomechanical measurements at average circulating photon numbers as low as $n_c = 0.27$, a new regime for diamond optomechanical devices, enabled by a high optomechanical coupling rate. 
Taking the likely origin of the measured mechanical damping rate of 3.36 kHz in Fig. \ref{fig:opto}b to be telegraph noise stemming from interactions with local two-level systems (TLS), this represents a mechanical coherence time of 160 $\upmu$s \cite{omalley_superconducting_2016}.
Future work involving both CW and pulsed measurements will explore sources of loss and decoherence in diamond mechanical resonators and may offer insight into surface imperfections that currently limit the effectiveness of quantum sensing schemes based on near-surface color centers \cite{myers_probing_2014}.
While further study of the interplay between these devices and their thermal environment in cryogenic conditions is required for operation in the quantum regime, our measurements illustrate the aptitude of single-crystal diamond as a host for optomechanical devices and constitute a first step towards utilizing diamond OMCs as intermediaries between quantum systems including both photons and spins. 

As a platform for hybrid quantum systems, one of the key attractive features of diamond is the potential for strong mechanical coupling to embedded spins, and we next discuss the prospects for both NV center and SiV center spins in diamond OMCs.
The spin–mechanical cooperativity is defined as $C_\mathrm{sm} = \frac{4g_\mathrm{sm}^2}{\gamma_\mathrm{m} \gamma_\mathrm{s}}$, where $g_\mathrm{sm}$ is the spin–mechanical coupling strength, and $\gamma_\mathrm{m}$ and $\gamma_\mathrm{s}$ are the mechanical and spin decay rates, respectively.
Using FEM simulations of the mechanical mode and reported stress/strain susceptibility of the NV center \cite{barfuss_spin-stress_2019}, we estimate $g_\mathrm{sm,NV}/2\pi \approx $ 132 Hz and $C_\mathrm{sm,NV} \approx $ 0.025 for an NV located in the high-strain region of our OMC device. 
With a longer T$_2$ enabled by dynamical decoupling at cryogenic temperatures (up to 0.6 s reported at 77 K \cite{bar-gill_solid-state_2013}), the expected $C_\mathrm{sm,NV}$ increases to $\approx $ 55.
The prospects are significantly improved for the SiV center which exhibits larger strain susceptibility than the NV center due to its strong strain-orbit coupling and different orbital character of the ground and excited spin qubit states \cite{meesala_strain_2018}.
For an SiV center, $g_\mathrm{sm, SiV}/2\pi$ reaches 8 MHz at 3.3 T and asymptotically approaches 9 MHz in the limit of high magnetic field, and the predicted $C_\mathrm{sm,SiV}$ exceeds 10$^9$ with the reported T$_2$ = 13 ms at 100 mK \cite{sukachev_silicon-vacancy_2017}.

\begin{backmatter}
\bmsection{Funding}
National Science Foundation (OMA-2016245, DMR-1906325, DMR-2308708, CNS-1725797, Award No. 2137740); The Netherlands Organisation for Scientific Research (024.003.037); UCSB Quantum Foundry; Eddleman Quantum Institute.

\bmsection{Acknowledgment}
We gratefully acknowledge support from the NSF QLCI program through Grant No. OMA-2016245, as well as the use of shared facilities of the UCSB Quantum Foundry through Q-AMASE-i program (NSF DMR-1906325), the UCSB MRSEC (NSF DMR-2308708), and the Center for Scientific Computing (CSC) (NSF CNS-1725797). 
This work was performed, in part, in the UCSB Nanofabrication Facility, an open access laboratory and at the Center for Integrated Nanotechnologies, U.S. Department of Energy's Office of Science User Facility jointly operated by Los Alamos and Sandia National Laboratories.
We particularly thank Demis John for help with the fabrication process development.
C. P., L. B. H, and  I. H. acknowledge support from the UCSB Quantum Foundry and H.O. and V. D. acknowledge support from the Eddleman Quantum Institute. D. B. acknowledges support from NSF Award No. 2137740 and NWO Grant No. 024.003.037.

\bmsection{Disclosures}
The authors declare no conflicts of interest.

\bmsection{Data availability} Data underlying the results presented in this paper are not publicly available at this time but may be obtained from the authors upon reasonable request.

\bmsection{Supplemental document}
See Supplement 1 for supporting content.
\end{backmatter}

\bibliography{references}

\end{document}


\maketitle

\section{Sample fabrication procedure}
\subsection{He implantation and subsurface graphitization}
An electronic-grade, (100)-oriented single-crystal diamond sample with dimensions of $2 \times 2 \times 0.5$ mm is obtained from Element Six.
The sample is polished and thinned to a thickness of 150 $\mu$m by Syntek, achieving a surface roughness of less than 300 pm.
Prior to ion implantation, the surface is etched several microns using \ch{Ar/Cl2} plasma (Oxford PlasmaPro 100 Cobra) to remove polishing-induced strain. Helium ions are implanted at 150 keV with a 7° incident angle to avoid channeling, with an implantation fluence of $5 \times 10^{16}$ / cm$^2$.
These parameters are selected based on Stopping and Range of Ions in Matter (SRIM) simulations, taking into account the damage threshold required for subsequent subsurface graphitization.
Following implantation, the sample is annealed at 850 °C for 8 hours in a high-vacuum chamber, resulting in the formation of a 100 nm-thick subsurface graphite layer beneath a partially damaged diamond layer approximately 400 nm thick. Between each process step, the sample is cleaned in a 1:1 mixture of \ch{H2SO4} and \ch{HNO3} at 160 °C for 2 hours.

\subsection{CVD diamond growth}
Diamond homoepitaxial growth and nitrogen doping are performed via plasma-enhanced chemical vapor deposition (PECVD) using a SEKI SDS6300 reactor.
The growth conditions consist of a 750 W plasma containing 0.4$\%$ $^{12}$CH$_{4}$ in 400 sccm H$_2$ flow held at 25 torr and $\sim$ 730 °C according to a pyrometer.
A 332 nm-thick isotopically purified (99.998$\%$ $^{12}$C) buffer layer is grown, followed by a 21 nm-thick $^{15}$N-doped layer (5 sccm $^{15}$N$_2$ gas, 10 minutes), and a 144 nm-thick $^{12}C$ capping layer. After growth, the sample is characterized with secondary ion mass spectrometry (SIMS) to estimate the isotopic purity, epilayer thickness, and properties of the nitrogen-doped layer.
The detailed conditions are previously reported in \cite{hughes_two-dimensional_2023}.
 
The diamond is further electron irradiated and annealed to generate enhanced NV center concentrations.
Irradiation is performed with the 200 keV electrons of a transmission electron microscope (TEM, ThermoFisher Talos F200X G2 TEM).
The sample then undergoes subsequent annealing at 850 °C for 8 hours in vacuum, during which the vacancies diffuse and form NV centers.
After irradiation and annealing, the sample is cleaned in a boiling triacid solution (1:1:1 H$_2$SO$_4$:HNO$_3$:HClO$_4$) and annealed in air at 450 °C for 4 hours to oxygen terminate the surface and help stabilize the negative NV$^{-}$ charge state for further measurements.

\begin{figure}[!htb]
\centering\includegraphics[width=0.70\textwidth]{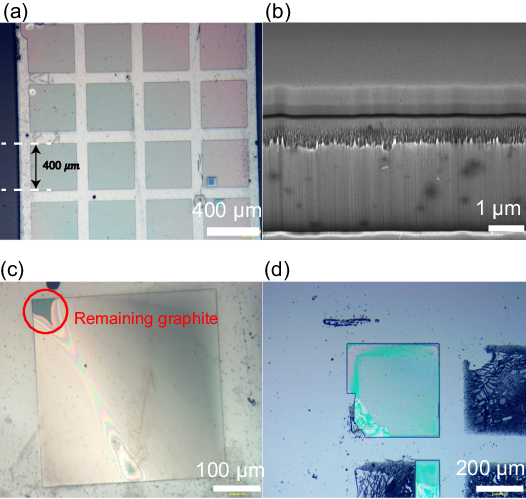}
\caption{\label{fig:app_fig1} 
Images during diamond membrane fabrication. (a) Microscope image of patterned square membranes on bulk diamond before electrochemical etching. (b) 45° SEM image of a single patterned membrane. The black layer denotes the subsurface graphite layer. (c) Microscope image of a diamond membrane after electrochemical etching and acid cleaning. The red circle indicates the remaining small graphite region below the diamond membrane for tethering. (d) Microscope image of transferred diamond membrane on a Si carrier piece.
}
\end{figure}

\subsection{Diamond membrane fabrication via smart-cut}
For robustness and ease of thin membrane transfer, small square membranes are patterned.
A 200 nm low-stress SiN layer is first deposited by PECVD (Plasmatherm Vision 310) to serve as a hard mask. A photoresist layer (AZ5214) is then spun on the SiN layer, followed by lithographic exposure using a maskless aligner (Heidelberg MLA150).
During this process, 16 square membranes of size 400 $\mu$m $\times$ 400 $\mu$m are defined.
The SiN layer is subsequently etched using a \ch{CHF3/O2} RIE (Panasonic E626I) to transfer the pattern.
After pattern transfer, the remaining photoresist is removed by soaking in NMP at 80 °C for 1 hour.
The diamond is then etched using \ch{O2} RIE (Panasonic E626I), and the residual SiN mask is removed by wet etching in o-phosphoric acid at 160 °C for 1 hour. (Fig. \ref{fig:app_fig1}a, b)

Following membrane patterning, the subsurface graphite layer is removed by electrochemical etching to enable membrane release.
The sample is immersed in a 250 mM \ch{K2SO4} solution, and a 30 V DC bias is applied between Pt wires (Thermo Scientific, 0.05 mm diameter) positioned near the target membrane.
The wire positions are adjusted using a micro-positioner in a probe station to localize etching to the desired region. (Fig. \ref{fig:app_fig1}c)
Etching proceeds until only a small tether of the graphite layer remains at a membrane corner, keeping the membrane attached to the bulk for controlled transfer.
The sample is then cleaned in a 1:1 mixture of \ch{H2SO4} and \ch{HNO3} to remove etching residues.

In parallel, a Si carrier piece of size $10 \times 10 \times 1$ mm is prepared by wafer dicing and subjected to the same acid cleaning.
An HSQ layer (Fox 16, Dow Corning) is spun on the Si carrier, and the diamond sample is flip-chip bonded onto it (Finetech Fineplacer Lambda).
The sample is flipped such that the grown side of the membrane is in contact with the HSQ.
The bonded stack is then placed into a graphite bonding fixture and pressed with a 400 N force, controlled by bolt torque.
The entire fixture is annealed in a tube furnace (Tystar 8300) under oxygen-free conditions at 450 °C for 8 hours to cure the HSQ, resulting in strong adhesion between the diamond membrane and the carrier substrate.

After curing, the diamond sample is carefully detached from the Si carrier using tweezers.
During this step, the target membrane, where the underlying graphite has been selectively removed, remains bonded to the carrier, while the other membranes are lifted away with the bulk diamond.
This process enables robust and repeatable transfer of multiple membranes from a single diamond chip.
The transferred membrane is immersed in buffered HF (BHF) for 1 minute to remove any exposed HSQ from the surface.
Finally, the membrane is thinned to the target thickness of 250 nm by \ch{Ar/Cl2} RIE, followed by \ch{O2} RIE surface termination (Panasonic E626I).
The membrane thickness is measured during and after etching by optical reflectometry (Filmetrics F40-UV).
Because the sample orientation is inverted during transfer, the partially damaged layer caused by He implantation is located above the pristine grown layer.
This damaged layer is fully removed during the final etching step. (Fig. \ref{fig:app_fig1}d)

\begin{figure}[!htb]
\centering\includegraphics[width=0.96\textwidth]{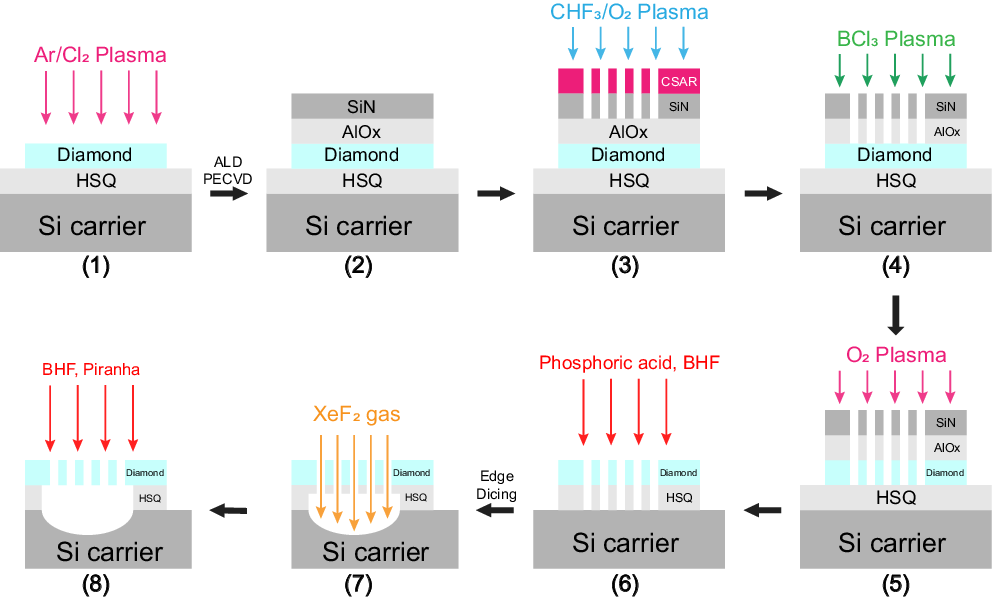}
\caption{\label{fig:app_fig2} 
Schematic of diamond OMC fabrication process after membrane transfer.
(1) Thin down diamond membrane t0 250 nm thickness via \ch{Ar/Cl2} RIE. (2) Deposit AlOx and SiN as mask materials. (3) Pattern OMC structure using EBL and etch SiN via \ch{CHF3/O2} RIE. (4) Etch AlOx via \ch{BCl3} RIE. (5) Etch diamond via \ch{O2} RIE. (6) Remove mask materials on diamond and exposed HSQ. (7) Undercut Si using \ch{XeF2} gas etching. (8) Remove HSQ at the backside of diamond and clean the sample.
}
\end{figure}

\subsection{OMC device fabrication}
The OMC device fabrication process is illustrated in Fig. \ref{fig:app_fig2}, which follows similar steps to membrane patterning, with additional layers and a different resist.
Since the OMC features are at the nanometer scale, electron-beam lithography (EBL) is employed for patterning.
As in the membrane process, etch mask layers are deposited on the transferred membrane sample.
A 25 nm AlOx layer is first deposited by atomic layer deposition (ALD, Oxford FlexAL), followed by a 200 nm low-stress SiN layer deposited by PECVD.
The additional AlOx layer serves as an etch-stop during SiN etching to prevent damage to the underlying diamond.
An electron-beam resist (CSAR62) is then spun onto the SiN layer and patterned using EBL (JEOL JBX-6300FS) to define the OMC structures.
The SiN layer is etched using \ch{CHF3/O2} RIE to transfer the pattern, and the remaining e-beam resist is removed by soaking in NMP at 80 °C for 1 hour.
The AlOx etch-stop layer is then removed using \ch{BCl3} RIE (Panasonic E626I), followed by diamond etching with \ch{O2} RIE.
The residual SiN mask is subsequently removed by wet etching in o-phosphoric acid at 160 °C for 1 hour.
The sample is then diced using a dicing saw (ADT 7100) within a few $\mu$m of the OMC device edge to allow optical access for lensed fiber coupling from the side.
After dicing, the sample is soaked in buffered HF (BHF) for 1 minute to remove any exposed HSQ.
To create a suspended OMC structure, the underlying silicon is isotropically undercut by a few $\mu$m using \ch{XeF2} gas etching (Xactix Xetch X3).
The sample is soaked again in BHF for 1 minute to remove HSQ from the backside of the diamond where the Si is etched.
Then the sample is annealed in air at 440 °C for 8 hours followed by a piranha clean (3:1 \ch{H2SO4}:\ch{H2O2}) for 2 hours.

\section{Material characterization}
\begin{figure}[!htb]
\centering\includegraphics[width=0.70\textwidth]{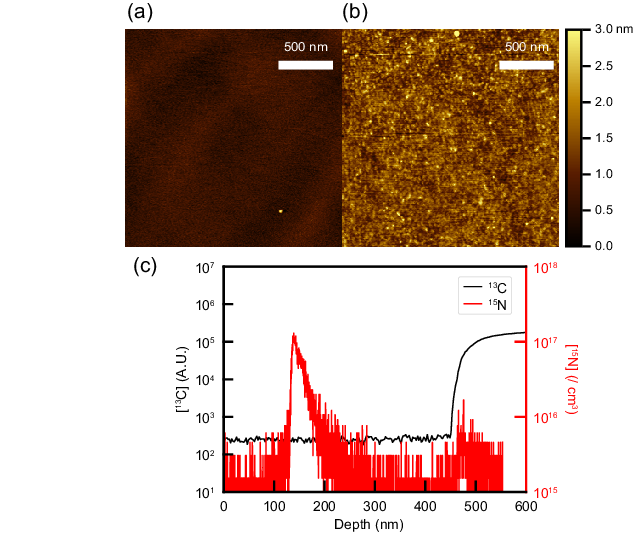}
\caption{\label{fig:app_fig3} 
Material characterization (a) AFM  image after diamond overgrowth. (b) AFM image after smart-cut membrane fabrication. (c) SIMS depth profile after diamond overgrowth. The black line indicates the $^{13}$C concentration, which shows the clear distinction between the purified grown layer and the original diamond with natural $^{13}$C abundance. The red line indicates the $^{15}$N concentration, showing confinement of $\delta$-doped layer in the middle.
}
\end{figure}

We characterize the surface roughness at key stages of the membrane fabrication process using atomic force microscopy (AFM) to monitor surface morphology, which is critical for achieving high-quality diamond OMC devices.
Fig. \ref{fig:app_fig3}a and \ref{fig:app_fig3}b show AFM images acquired after diamond overgrowth and after membrane fabrication, respectively.
Following diamond growth (Fig. \ref{fig:app_fig3}a), the root-mean-square surface roughness is measured to be $R_q = 139.9$ pm, which is comparable to the initial roughness observed after fine polishing.
After membrane fabrication (Fig. \ref{fig:app_fig3}b), the surface roughness increases slightly to $R_q = 376.7$ pm, yet remains smoother than that of a new, unpolished diamond sample.
This moderate increase in roughness is attributed to the \ch{O2} RIE surface termination step.

We also characterize the diamond overgrowth and $^{15}$N $\delta$-doping using secondary ion mass spectrometry (SIMS).
Fig. \ref{fig:app_fig3}c shows the SIMS depth profiles of $^{13}$C and $^{15}$N concentrations.
The $^{13}$C profile reflects the distinction of isotopically purified diamond overgrowth, exhibiting a lower $^{13}$C concentration compared to natural abundance.
From the sharp increase in $^{13}$C concentration, we estimate the overgrown layer thickness to be approximately 497 nm.
A distinct peak in $^{15}$N concentration at a depth of 144 nm indicates a confined 2D $^{15}$N layer embedded within the grown diamond.
From this profile, the $^{15}$N density is estimated to be 7.5 ppm $\cdot$ nm.

\section{OMC design optimization}
The optomechanical coupling strength and quality factors of an OMC device strongly depend on design parameters such as unit cell dimensions and hole shapes.
Therefore, optimization of the OMC geometry is essential.
We follow the parameterization strategy of the full OMC structure described in \cite{chan_laser_2012}.
Based on this parameterization, both optical and mechanical modes are simulated using COMSOL Multiphysics with the finite element method (FEM).
These simulations yield the optical and mechanical mode frequencies ($\omega_\mathrm{o}$, $\omega_\mathrm{m}$), the optomechanical coupling strength ($g_0$), and the optical quality factor ($Q_\mathrm{o}$).

To reduce the dimensionality and complexity of the optimization problem, the process is divided into two stages.
The first stage focuses on mirror cell optimization and the second stage addresses the deformation profile optimization.
In the first stage, the geometry of a single mirror cell is optimized to maximize both the optical and mechanical bandgaps.
Given the small number of parameters involved, this step is performed using a grid search.
After determining the optimal mirror cell parameters, we proceed to optimize the deformation profile from the mirror cells toward the defect region.
Since this parameter space is larger and more complex, we apply a genetic algorithm to maximize a fitness function defined as

\begin{equation}
    \mathcal{F} = g_0 Q_\mathrm{o} \cdot G(\omega;\mu_\mathrm{o}, \Delta_\mathrm{o}/6) \cdot G(\omega;\mu_\mathrm{m}, \Delta_\mathrm{m}/6)
\end{equation}
\\
where $G(\omega;\mu, \sigma) = \exp\left( -\frac{(\omega - \mu)^2}{2\sigma^2} \right)$ represents a Gaussian function centered at $\mu$ with standard deviation $\sigma$.
The parameters $\mu_\mathrm{o}$ and $\mu_\mathrm{m}$ denote the center frequencies of the optical and mechanical bandgaps, respectively, and $\Delta_\mathrm{o}$ and $\Delta_\mathrm{m}$ represent their corresponding bandwidths.
The product $g_0 Q_\mathrm{o}$ promotes designs with both strong optomechanical coupling and high optical quality.
The Gaussian terms favor optical and mechanical mode frequencies located near the center of their respective bandgaps.
The standard deviation $\sigma = \Delta / 6$ ensures that the $3\sigma$ range of each Gaussian coincides with the edges of the corresponding bandgap.

The mode simulations are performed using Python-wrapped COMSOL codes, and the optimization routines are implemented in Python. 
The genetic algorithm is carried out using the PyGAD package.
To run the genetic algorithm with parallelization, a High-performance computing (HPC) cluster is used.

\section{Optomechanics characterization}
\subsection{Measurement setup}
\begin{figure}[!htb]
\centering\includegraphics[width=0.96\textwidth]{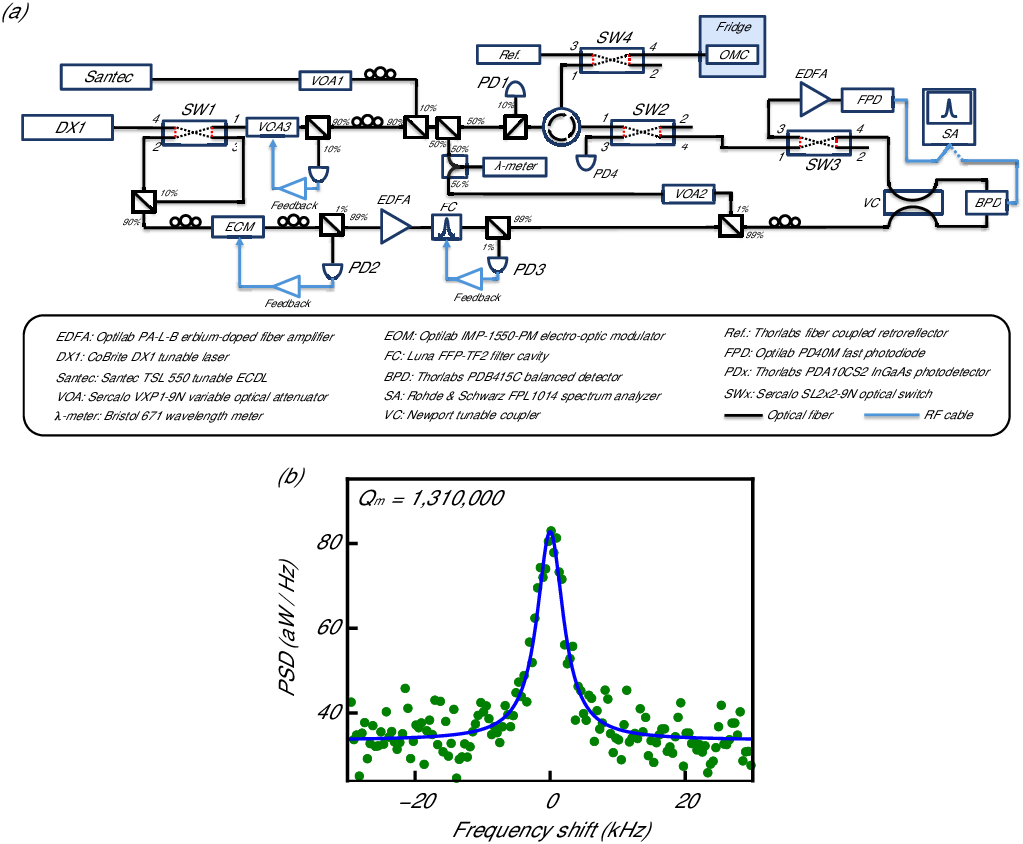}
\caption{\label{fig:app_fig4} 
Optomechanics characterization. (a) Detailed schematic of optomechanics measurement setup. (b) Mechanical resonance of the diamond OMC at T = 160 mK. The blue line is a Lorentzian fit of the spectrum, which gives a mechanical quality factor of $(1.31 \pm 0.05) \times 10^6$.
}
\end{figure}

Fig. \ref{fig:app_fig4}a shows the complete measurement setup used to perform optical and mechanical characterization. Light from either one of two lasers is sent to the OMC via a circulator to probe the optical and mechanical modes of the OMC. Two variable optical attenuators (VOAs), VOA1 and VOA3, allow us to switch between the Santec laser and DX1 lasers. The piezo tuning of the Santec laser allows for much faster wavelength sweeping for optical characterization compared to the thermal tuning of the DX1 laser. The reflected light, which contains photons upconverted or downcoverted by the mechanics as well as pump light that did not interact with the mechanics, is routed to towards the detection chain by the circulator. The powers measured at photodiodes PD1 and PD4, along with the measured insertion losses of the optical elements in between, are used to measure the coupling efficiency between the lensed fiber and the waveguide on the device. This coupling efficiency, along with the input power calculated by measuring the power incident at PD1, is used to calculate the average number of intracavity photons ($n_\mathrm{c}$) in the OMC using the equation
\begin{equation}\label{eqn:nc}
    n_\mathrm{c} = \frac{\kappa_\mathrm{e}}{\Delta^2 + \frac{\kappa^2}{4}} \cdot \frac{P_{\mathrm{in}}\eta_{\mathrm{f}}}{\hbar \omega_\mathrm{L}},
\end{equation}
where $\eta_{\mathrm{f}}$ is the fiber coupling efficiency, $P_{\mathrm{in}}$ is the optical power in the fiber going to the device, $\kappa_\mathrm{e}$ is the input coupling rate from the waveguide to the cavity, and $\Delta = \omega_\mathrm{L} - \omega_0$ is the detuning of the laser from the cavity resonance. For mechanical characterization, the setup allows to perform simple heterodyne measurements as well as balanced heterodyne measurements. Simple heterodyne measurements are used for obtaining the mechanical power spectral density at high $n_\mathrm{c}$ (typically $>$200), and the more sensitive balanced heterodyne measurements are used for intracavity photon numbers as low as 0.25.
For all measurements, we use a Bristol 871 wavemeter to accurately measure the laser wavelength. Manual fiber polarization controllers are used for each laser to adjust the polarization of the light to maximize the fiber-to-waveguide coupling.

\subsection{Optical characterization}
To extract the resonance wavelength and quality factor of the optical mode of the OMC, the laser wavelength is monitored by a wavelength meter as it is swept over the resonance, and the reflected power as a function of frequency is fit to the equation
\begin{equation}\label{eqn:reflection}
    R(\Delta) = y_0 - a_0 \cdot \frac{(1 - q^2)\kappa/2 - q\Delta}{\kappa^2/4 + \Delta^2},
\end{equation}
where $\omega_\mathrm{L}$ is the laser frequency, $\omega_\mathrm{o}$ is the frequency of the optical resonance, $\Delta = \omega_\mathrm{L} - \omega_\mathrm{o}$ is the detuning of the laser from resonance, $\kappa$ is the linewidth of the optical mode, and $q$ is the Fano parameter.
Eq. \ref{eqn:reflection} accounts for a Fano-like optical response, likely originating from interference between light reflected from the lensed fiber tip, the waveguide tip, and the mirror patterned at the end of the waveguide \cite{tamaki_two-dimensional_2024}.
The fixed polarization condition of the waveguide imparts polarization dependence to the reflected light as well.
This effect has been modeled in \cite{tamaki_two-dimensional_2024} as an effective beamsplitter with a polarization-dependent reflectivity. 
The coupling of the optical mode to the waveguide $\kappa_{e}$ is estimated from the contrast of the optical resonance dip and verified by a separate measurement of the amplitude and phase response of the optical mode.

\subsection{Mechanical characterization}
For mechanical power spectral density measurements, we use both simple heterodyne and balanced heterodyne measurement schemes. For simple heterodyne measurements, a GHz-bandwidth fast photodiode (FPD) is used to detect a beat note at the mechanical frequency $\omega_\mathrm{m}$ between the reflected pump laser tone and the optomechanically upconverted or downconverted photons. 

At lower $n_\mathrm{c}$, where the main laser pump tone and the optomechanically generated tone are both much smaller, we use a balanced heterodyne measurement scheme that offers a much higher sensitivity.
We capture the beating of the optomechanically generated photons with a local oscillator (LO) tone produced by splitting off 90\% of the laser power, amplifying it with an L-band erbium-doped fiber amplifier (EDFA: Optilab PA-L-B), and frequency shifting it using an electro-optic modulator (EOM: Optilab IMP-1550-PM). The local oscillator frequency is set to be 50 MHz away from the mechanically up- or down-converted tone. This brings the beating tone within the bandwidth of the balanced photodetector (BPD: Thorlabs PDB415C). The local oscillator and the optomechanically-generated tones are mixed and balanced using a variable coupler (Newport tunable coupler). To maximize LO power without saturating the detector, we filter out EDFA amplified spontaneous emission (ASE) and the main laser tone by locking a tunable filter cavity to the EOM-generated LO tone. VOA2 and VOA3 are used to control which laser tones are measured at the wavemeter, allowing us to independently measure the wavelengths of the pump and LO tones. The polarization controller on the LO arm is used to match the polarization of the LO to the signal tone to maximize the amplitude of the RF beat note after photodetection. The RF signal from either the BPD or the FPD is captured using a spectrum analyzer (SA: Rohde and Schwarz FPL1014). 

The measurement of mechanical quality factor $Q_\mathrm{m}$ shown in the Main Text Fig. 3b is performed at a sample stage temperature of 4 K.
We also perform a similar measurement at a sample stage temperature of 160 mK in the dilution refrigerator.
Fig. \ref{fig:app_fig4}b shows the measured power spectral density for such a measurement at $n_\mathrm{c} = 0.5$.
The mechanical mode frequency is $\omega_\mathrm{m}/2\pi = 6.23$ GHz, and the Lorentzian fit yields $Q_\mathrm{m} = (1.31 \pm 0.05) \times 10^6 $.

\subsection{Summary of device parameters}

The diamond smart-cut method enables robust and scalable fabrication of nanomechanical devices.
Compared to the previous approach based on the DOI method \cite{cady_diamond_2019}, the device yield is improved, and the fabrication cycle time is significantly reduced.
Table \ref{tab:app_tab1} summarizes the measured optical and mechanical properties from several devices on the same chip.
Device A is the device presented in the Main Text.

\begin{table}[!htb]
\centering
\begin{tabular}{|c|c|c|c|c|c|}
\hline
Device & $\lambda_\mathrm{o}$ (nm) & $\kappa / 2\pi$ (GHz) & $\kappa_\mathrm{e}/2\pi$ (GHz) & $\omega_\mathrm{m} / 2\pi$ (GHz) & $\gamma_\mathrm{m} / 2\pi$ (kHz) \\
\hline
Device A & 1576.87 & 5.03 & 2.99 & 6.23 & 3.28\\
Device B & 1573.28 & 4.29 & 0.55 & 6.24 & 37.3\\
Device C & 1569.44 & 4.57 & 0.58 & 6.22 & 42.8\\
Device D & 1567.05 & 2.15 & 0.41 & 6.22 & 11.6\\
Device E & 1566.77 & 5.30 & 1.59 & 6.18 & 3.79\\
\hline
\end{tabular}
\caption{\label{tab:app_tab1}Measured optical and mechanical device parameters. This table shows measured optical and mechanical properties from different OMC devices on the same chip. $\lambda_\mathrm{o}$ is optical resonance wavelength, $\kappa$ is measured optical damping rate, $\omega_\mathrm{m}$ is mechanical resonance frequency, and $\gamma_m$ is measured mechanical damping rate.}
\end{table}

\subsection{Comparison of mechanical resonator properties}
Table \ref{tab:app_tab2} summarizes the reported mechanical properties from OMC devices in various materials under CW laser measurement at cryogenic temperatures. We include the $Qf$ product for these devices, a figure of merit that quantifies the decoupling of the mechanical resonator from its thermal environment and from parasitic impurities such as two-level systems (TLS). This high $Qf$ product implies a low thermal decoherence rate.

\begin{table}[!htb]
\centering
\begin{tabular}{|c|c|c|c|c|}
\hline
Resonator Type & Quality Factor & Mode frequency (GHz) & $Qf$ Product (Hz) & Reference \\
\hline
GaAs OMC & $1.1 \times 10^3$ & 2.37 & $2.6 \times 10^{12}$& \cite{ramp_elimination_2019}\\
LN OMC & $1.7 \times 10^4$ & 1.9 & $3.2 \times 10^{13}$    & \cite{jiang_lithium_2019}\\
GaP OMC & $2.11 \times 10^5$ & 2.91 & $6.12 \times 10^{14}$ & \cite{stockill_gallium_2019}\\
Diamond OMC & $4.4 \times 10^5$ & 5.76 & $2.53 \times 10^{15}$ & \cite{joe_high_2024} \\
Si OMC (1D) & $1.31 \times 10^6$ & 5.31 & $6.96 \times 10^{15}$ & \cite{maccabe_nano-acoustic_2020} \\
Si OMC (2D) & $7.02 \times 10^5$ & 10.21 & $7.16 \times 10^{15}$ & \cite{ren_two-dimensional_2020} \\
Diamond OMC & $1.90 \times 10^6$ & 6.23 & $1.18 \times 10^{16}$  & This work \\
\hline
\end{tabular}
\caption{\label{tab:app_tab2} Comparison of mechanical resonator properties under continuous-wave (CW) illumination. This table shows the mechanical properties of optomechanical crystal devices fabricated from diamond and other materials, measured at cryogenic temperatures. We note that under pulsed optical probing, higher $Qf$ products have been measured in Si OMCs.}
\label{tab:resonators}
\end{table}

\bibliography{references}